%% file: main.tex
\documentclass{ar-1col-S2O}

\usepackage{amsfonts,amsmath,amssymb,amsthm}
\usepackage{booktabs}
\usepackage{longtable}
\usepackage{float}
\usepackage{lipsum}
\usepackage[ruled,vlined]{algorithm2e}
\usepackage{caption}
\usepackage{subcaption}
\usepackage[acronym]{glossaries}
\usepackage[backgroundcolor=white]{todonotes}

\usepackage{amssymb}
\usepackage{pifont}
\newcommand{\cmark}{\ding{51}}%
\usepackage{multirow}
\usepackage{tabularx}
\theoremstyle{definition}

\theoremstyle{remark}

\usepackage{array} 

\theoremstyle{plain}

\theoremstyle{definition}

\usepackage[comma]{natbib}
\usepackage{url}
\jname{Xxxx. Xxx. Xxx. Xxx.}
\jvol{AA}
\jyear{YYYY}
\doi{10.1146/((please add article doi))}

\begin{document}

\markboth{Bai et al.}{RL in Finance}

\title{A Review of Reinforcement Learning in Financial Applications}

\author{Yahui Bai\textsuperscript{*}\thanks{* Equal contribution, order by alphabet.},$^1$ Yuhe Gao\textsuperscript{*},$^1$ Runzhe Wan\textsuperscript{*},$^2$ Sheng Zhang\textsuperscript{*},$^2$ and Rui Song\textsuperscript{\dag}$^2$ \thanks{\dag 
  \ Corresponding author.}
\affil{$^1$Department of Statistics, North Carolina State University, Raleigh, USA; email: byahui1@gmail.com, ygao32@ncsu.edu}
\affil{$^2$Amazon, Seattle, USA; email: runzhe.wan@gmail.com, shengzh01@gmail.com, songray@gmail.com}}

\begin{abstract}
In recent years, there has been a growing trend of applying Reinforcement Learning (RL) in financial applications. 
    This approach has shown great potential to solve decision-making tasks in finance.
    In this survey, we present a comprehensive study of the applications of RL in finance and conduct a series of meta-analyses to investigate the common themes in the literature, 
    such as the factors that most significantly affect RL's performance compared to traditional methods.
    Moreover, we identify challenges including explainability, Markov Decision Process (MDP) modeling, and robustness that hinder the broader utilization of RL in the financial industry and discuss recent advancements in overcoming these challenges. 
    Finally, we propose future research directions, such as benchmarking, contextual RL, multi-agent RL, and model-based RL to address these challenges and to further enhance the implementation of RL in finance.
\end{abstract}


\maketitle

\tableofcontents


\input{1_Intro}

\input{2_Pre}

\section{Applications of RL in Finance}\label{sec: application}

\input{3_0_tables.tex}

\input{3_2_MM}

\input{3_3_PM}
\input{3_4_OE}
\input{4_Meta_Analysis}

\input{6_env}

\input{5_challenges}
\input{6_Future_Directions}






%




\bibliographystyle{ar-style1.bst}
\bibliography{ref1}





\newpage

\appendix

\input{7_appendix}



\end{document}

%% file: 1_Intro.tex
\section{Introduction}%
\label{sec:intro}

A financial market is a marketplace where financial instruments such as stocks and bonds are bought and sold \citep{fama1970efficient}. 
Individuals and organizations can play crucial roles in financial markets to facilitate the allocation of capital. 
Market participants face diverse challenges, such as portfolio management, which aims to maximize investment returns over time, and market-making, which seeks to profit from the bid-ask spread while managing inventory risk.
As the volume of financial data has increased dramatically over time, new opportunities and challenges have arisen in the analysis process, leading to the increased adoption of advanced Machine Learning (ML) models. 

Reinforcement Learning (RL)\citep{sutton2018reinforcement}, as one of the main categories of ML, has revolutionized the field of artificial intelligence by empowering agents to interact with the environment and allowing them to learn and improve their performance. 
The success of RL has been demonstrated in various fields, including games, robots, mobile health \citep{nash1950equilibrium, kalman1960new, Murphy2003}, etc.
In finance, applications such as market making, portfolio management, and order execution can benefit from the ability of RL algorithms to learn and adapt to changing environments. 
Compared to traditional models that rely on statistical techniques and econometric methods such as time series models (ARMA, ARIMA), factor models, and panel models, the RL framework empowers agents to learn decision-making by interacting with an environment and deducing the consequences of past actions to maximize cumulative rewards \citep{charpentier2021reinforcement}. 
RL also facilitates online learning, allowing iterative refinement of investment strategies with new information. 
Moreover, through its deep network structure, RL models can better capture complex patterns in highly nonlinear and non-stationary financial data. 
These advantages give RL the potential to enhance both the efficiency and effectiveness of financial applications.


Despite numerous studies that demonstrate performance enhancements compared to conventional methods, it is crucial to thoroughly assess the challenges associated with applying RL in the financial domain. 
Financial data can be noisy and non-stationary, while often follows a heavy-tail distribution, and may involve a mix of frequencies.
These features can present challenges to RL algorithms both in theory and practice.
This survey aims to provide an overview of recent studies on the use of RL in three critical finance applications: Market Making, Portfolio Management, and Optimal Execution. The main focus is to highlight challenging areas where further exploration would be potentially valuable.
For all three areas mentioned above, we select recently published papers with a focus on those with high-quality and relevant experiments that can support our meta-analysis.

\subsection{Related Work}



Several papers have reviewed the use of RL in finance. 
\citet{fischer2018reinforcement} and \citet{pricope2021deep} categorize model-free RL approaches into critic-only, actor-only, and actor-critic, and discuss various RL settings. 
Specifically, \citet{fischer2018reinforcement} summarized the challenges in financial data and suggested the actor-only approach may be the best-suited for RL in financial markets. 
\citet{pricope2021deep} identifies limitations in current studies on deep reinforcement learning (DRL) in quantitative algorithmic trading and suggests that more research is needed to determine its ability to outperform human traders. 
\citet{charpentier2021reinforcement} and \citet{huang2020deep} delve deep into the realm of RL applications, with a primary focus on their widespread use in economics, operations research, and banking.
Other works focus on application in subdomain of finance. \citet{hambly2021recent} conducts a survey of recent studies in RL applied to finance, concentrating on critical topics like optimal execution and portfolio optimization,
while \citet{gavsperov2021reinforcement} focuses on the review of RL in market making.
\citet{meng2019reinforcement} diligently reviews research papers related to trading and forecasting in stock and foreign currency markets, shedding light on the prevalence of unrealistic assumptions within many of these studies,
and \citet{millea2021deep} reviews recent developments in DRL for trading in the cryptocurrency market.
\citet{mosavi2020comprehensive} compares ML and DRL methods in different financial applications, suggesting DRL may outperform traditional approaches.

Traditionally, RL methods have been applied to healthcare, such as precision medicine \citep{chakraborty2014dynamic, kosorok2019precision}. Key differences between RL in finance and medical research include: many RL agents in precision medicine are trained offline using observational or clinical trial data, while most RL policies in finance are trained online with simulators or real market data; traditional precision medicine settings typically involves finite horizons, whereas RL in financial markets often deals with infinite horizon decision making; financial environments can involve multiple agents, such as in market making, while precision medicine usually involves a single RL agent. It is worth to mention that, RL applications in mobile health devices are similar to those in finance, as both involve online RL with infinite decision stages.

\subsection{Contribution}

In this paper, we conduct a comprehensive survey of RL methods in the financial domain. 
Our contributions are summarized as follows. 
\begin{itemize}
    \item We provide a few taxonomies to establish a unified view of the recent literature about RL in financial applications including market making with a single agent and multiple agents, portfolio management, and optimal execution.
    \item We conduct a series of meta-analyses based on the reviewed papers to investigate the common patterns in the literature, such as which factor affects the performance of RL the most compared to traditional methods.
    \item We outline three key challenges that emerge when using RL in financial applications. In addition, we dive into a comprehensive discussion on the recent advances and progress made in addressing these challenges. 
    We also suggest six future directions for these problems.
\end{itemize}

\noindent The rest of the paper is organized as follows. Section \ref{sec:preliminary} summarizes the concepts of RL and commonly used RL algorithms in financial applications.
Section \ref{sec: application} provides a collection of RL papers in financial domains. 
Section \ref{sec:meta-analysis} presents meta-analysis in different papers to better understand the performance of RL in financial applications.
Sections \ref{sec: challenge} discusses challenges for the application of RL in the financial domain. 
Sections \ref{sec: data discussion} discusses the current status of environments, benchmaking and open-sourced packages in finance RL. 
\ref{sec: future}  concludes with a few future research directions.

%% file: 2_Pre.tex
\section{Preliminary} \label{sec:preliminary}

\begin{figure}
    \centering
    \includegraphics[width=1\textwidth]{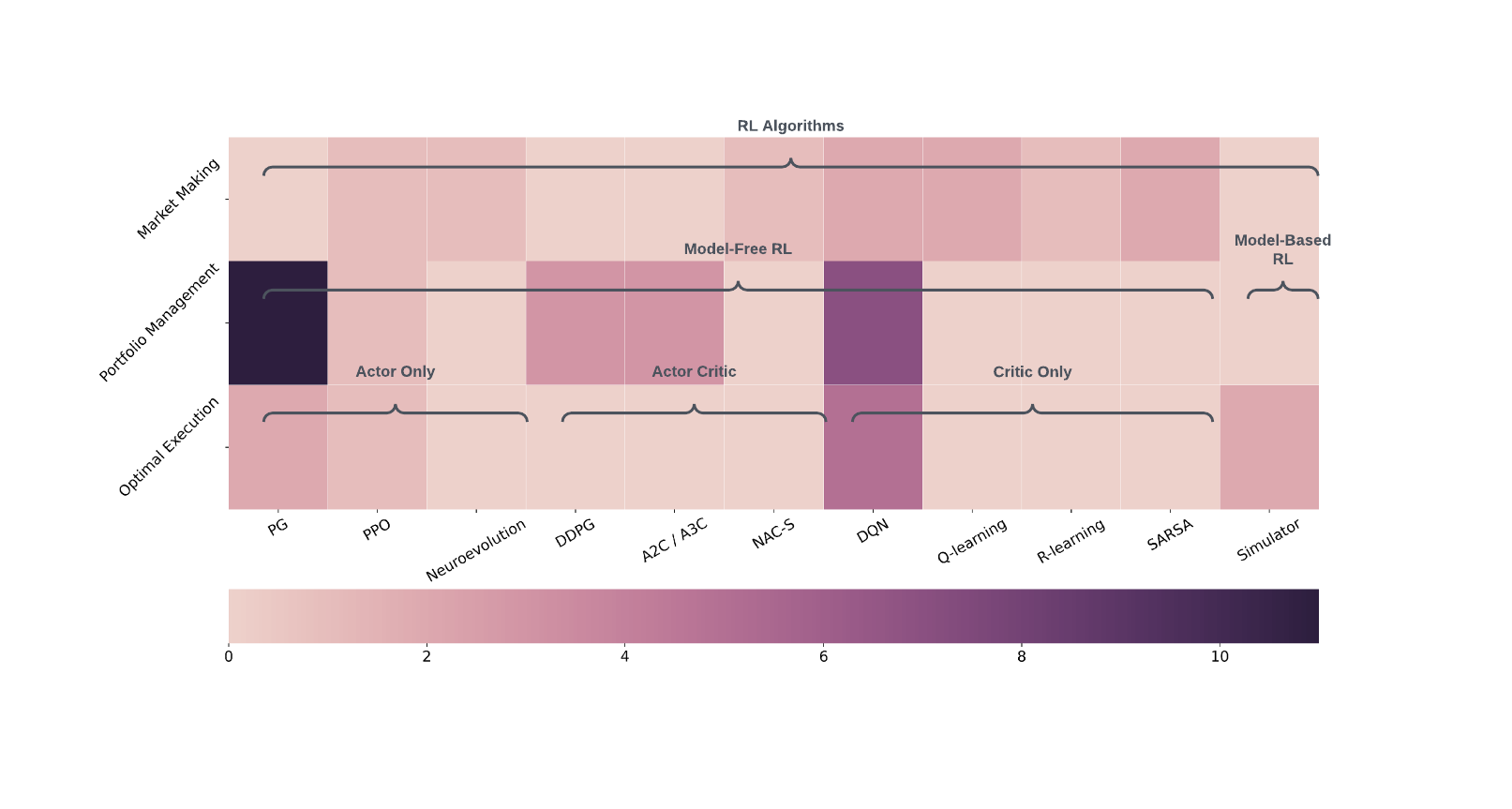}
    \vspace{-30pt}
    \caption{Frequency of the RL algorithms used in the literature.}
    \label{fig:method}
\end{figure}

In our financial literature survey, we introduce RL algorithms, categorized by frequency in Figure \ref{fig:method} into model-free and model-based algorithms. 
In model-free RL algorithms, the agent learns to make decisions without a model of the environment's dynamics, while in model-based RL algorithms, the environment's dynamics is modeled.

There are three categories of model-free RL algorithms, including actor-only, critic-only, and actor-critic methods.
Actor-only algorithms update only the policy, or the actor, while keeping the value function fixed.
One popular actor-only algorithm is called the Policy Gradient (PG) \citep{sutton1999policy}, which directly optimizes the policy parameters to maximize the expected cumulative reward. 
Another commonly used actor-only algorithm is Proximal Policy Optimization (PPO).
One of the key features of PPO is that it employs a ``proximal" update rule, which limits the size of the policy update at each iteration. 
Neuroevolution \citep{such2017deep} differs from the aforementioned algorithms as it adapts a gradient-free genetic algorithm to optimize the policy.  

Critic-only methods focus solely on learning the value function, without explicitly learning a policy function. The actions are obtained based on the approximate value function  \citep{sutton2018reinforcement}.
Q-learning estimates the optimal action-value function, which represents the expected total reward for taking a given action in a given state and following the optimal policy thereafter \citep{watkins1992q}.
Deep Q-Network (DQN) \citep{mnih2013playing} uses a deep neural network to estimate the Q-value function. 
Another type of critic-only algorithm is called SARSA \citep{sutton2018reinforcement}, which updates the Q-value function using the observed reward and the next state-action pair.
Additionally, R-learning \citep{schwartz1993reinforcement} is also value-based, except that its objective is to maximize the average reward rather than the discounted reward in Q-learning and SARSA. 

Actor-critic methods combine the benefits of both policy-based (actor) and value-based (critic) approaches.
These methods leverage two separate components to learn both the policy and value functions, with the actor generating actions based on the learned policy and the critic network evaluating the quality of the actions taken by the actor. 
Advantage Actor-Critic (A2C) and Asynchronous Advantage Actor-Critic (A3C) are popular and effective actor-critic methods with deep learning techniques.
A2C proposed by \cite{mnih2016asynchronous} combines the actor-critic algorithm with parallelization techniques to improve the learning speed and stability.
As an extension of A2C, A3C employs asynchronous updating of the actor-critic model by running multiple threads in parallel.
Deep Deterministic Policy Gradient (DDPG) \citep{lillicrap2015continuous} combines ideas from the deterministic policy gradient algorithm and the DQN algorithm, which is particularly effective in continuous action spaces.
The NAC-S($\gamma$) algorithm, proposed in \citet{thomas2014bias}, is a natural actor-critic method that uses semi-gradient SARSA for policy evaluation and is specifically designed for stochastic policies.

Model-based methods
involve learning a model of the environment's dynamics and using this model to simulate the future to select actions.
In financial applications like market making and optimal execution, the model-based RL approach entails constructing and leveraging a simulator that mimics the dynamics of the financial market. This simulator acts as a predictive model, providing agents with valuable insights about how the market may behave under different conditions and scenarios. By employing this model as a proxy of the real market, agents can execute a multitude of trades and test various trading strategies.
While these methods can be computationally intensive, they are shown to produce better long-term behavior when the environment is complex and uncertain \citep{sutton2018reinforcement}.

%% file: 3_0_tables.tex
\subsection{Summary of MDP Formulations in the Literature}

We first present a taxonomy of the state, action, reward, and algorithm used in the surveyed papers, to provide a unified view of the literature in Table \ref{table:state_action} and \ref{table:reward_algo} in Appendix \ref{sec:appendix_ref_table}. 
We organize the state, action and reward variables into several categories: (1) The state variables include 
the price information  (historical prices, returns, change of price, etc.), inventory information (the on-hand units for one stock or every stock in the portfolio), technical indicators (Relative Strength Index, Moving Average Convergence Divergence, etc., denoted as ``tech-indicators in Table \ref{table:state_action}''), market indicators (Citigroup Economic Surprise Index, etc.), company information (P/E ratio, dividend, market capitalization, etc.), predictive representations of state, trading volume (including volume of past trades, change of volume, market share of different agents, etc.),  bid-ask information (bid-ask spread, imbalance of bid-ask order in the limit order book, etc.), volatility, time index, and end-to-end information (E2E, where we directly use a large number of observable variables). 
(2) The action variables are summarized as long/short/hold, the target portfolio weight, the bid price, and the trading quantity. 
(3) The reward variables are summarized as slippage, implementation shortfall (IS), PnL (including cash income and outflow, and some other customized function of profit), return, volatility-adjusted return (VR, which includes the Sharp ratio and other variants), inventory penalty, market quality (containing bid-ask spread, some measures of price change, etc, as discussed in \citet{chan2001electronic}), and various shaped rewards. Here, a \textit{shaped reward} \citep{lin2020deep, lin2020E2E} refers to a reward definition that is not the same as any existing metrics but instead has a relatively complicated (but carefully designed) structure to achieve a special goal. 




%% file: 3_2_MM.tex
\subsection{Applications in Market Making}\label{sec:mm}

Market makers increase the liquidity of the market by continuously offering buy and sell orders. In general,  market makers profit from order execution (i.e., the \textit{spread profit}) as well as inventory \textit{speculation} resulting from changes in the market prices of the assets they hold.
However, maintaining a non-zero inventory also exposes market makers to the \textit{inventory risk}, which is the potential risk of adverse price change in the market \citep{o1986microeconomics}. 
Thus, market making algorithms aim at maximizing profit while controlling this inventory risk. RL methods have been introduced to market making since 2001 \citep[see][]{chan2001electronic}, with most works emerging in the last five years (post-2018). 
Most researchers use a single RL agent to handle the problem of market making while incorporating techniques to mitigate various risks, 
a representative selection of which is reviewed in Section \ref{sec:single_agent_mm}. Alternatively, some works investigate a multi-agent framework to achieve various objectives, such as 
increasing the robustness of the RL agent from some adversaries or 
 deriving a nearly optimal strategy in the presence of competing market makers.  An in-depth review of these approaches is available in Section \ref{sec: multi_agent_MM}. 


\subsubsection{Single-agent RL in Market Making}\label{sec:single_agent_mm}

The application of Reinforcement Learning (RL) in market making is first introduced by \citep{chan2001electronic}. This pioneering work integrates the problem of market making into the RL framework, considering both the agent's information (such as inventory) and market characteristics (like the imbalance of buy and sell orders, bid-ask spread, and price changes between trades) as state variables. The RL agent's actions range from setting bid/ask prices, setting bid/ask order sizes, and executing buy/sell orders. The rewards of RL agents come from profit maximization, inventory minimization, and improvement of market quality. In the simulation model of \citep{chan2001electronic}, a single security is traded and state variables are influenced by the actions of the market maker, informed traders (with inside information distribution), and uninformed traders.

Subsequent research focuses on mitigating inventory and other risks for the RL market maker agent in various settings. \citet{mani2019applications} introduces a novel approach using the double SARSA algorithm and a customized temporal difference update for the Q-function. This approach leads to variance reduction in rewards, enabling the agent to achieve higher profits while maintaining lower inventory levels compared to the standard SARSA policy from \citep{chan2001electronic}. Another line of research involves modifying the reward function to manage risks. For instance, \citet{Spooner2018MM} designs an asymmetrically dampened reward function that encourages the agent to receive rewards by providing more liquidity rather than holding a high inventory, thereby reducing the inventory risk.  \citet{selser2021optimal} proposes to modify the reward function by incorporating the estimated variance of the agent’s wealth as a penalty term. In the simulation results of \cite{selser2021optimal}, it is demonstrated that a DQN model with this reward function can outperform a non-RL baseline that solves partial differential equations \citep{avellaneda2008high}. \citet{gueant2019deep} extends this risk-adverse reward function strategy to a multi-asset market making setting. They propose an actor-critic algorithm to approximate the optimal quote for multiple bonds.
Their RL model includes a penalty term for the portfolio's PnL variance in the reward function, and this approach is proved to be scalable in high-dimensional cases involving up to 100 bonds – a task challenging for non-RL methods.

\subsubsection{Multi-agent RL in Market Making}\label{sec: multi_agent_MM}



The multi-agent RL framework is an extension to the single-agent RL framework for market making, and the research directions can be divided into three branches. The first branch introduces adversarial agents in simulations, such as these agents profit from the losses of RL market-making agents. This approach enhances the RL agent's robustness to market conditions or increases risk aversion. For example, \citet{spooner2020robust} explore an adversarial RL setting in which an adversarial agent during the RL agent's training alters the environmental parameters in simulators. This discrepancy in perceived versus actual market conditions trains the market maker to be more robust against market condition misspecifications. Extending this idea, \citet{gavsperov2021market} introduces an adversary that directly disrupts the actions of the market making agent by displacing quotes. This technique has shown to reduce inventory risk and improve profits compared to training the RL agent without an adversary.

The second branch focuses on competition between market makers, where a professional market making agent competes with the RL agent. \citet{Ganesh2019MM} first demonstrate theoretically that the RL market making agent's optimal pricing policy to maximize spread PnL depends on its competitors' pricing distributions. 
\citet{Ganesh2019MM}  show in their experiments that
the RL agent can learn the pricing distribution of competing market makers from pricing and trading observations and derive a nearly optimal strategy accordingly, despite that the parameters of the competitor agents are  unknown to the RL agent.   In the third branch, the focus is mainly on adapting the original market making decision process into a hierarchical one. \citet{Patel2018MM} transform the RL framework by introducing a macro-agent that determines the overall trading strategy (buy, sell, or hold the security), while a micro-agent sets specific quote prices based on the macro-agent's direction. This hierarchical approach results in less volatile profits, as shown in their simulations.

%% file: 3_3_PM.tex
\subsection{Applications in Portfolio Management}\label{sec:Portfolio Management}

In portfolio management, the agent aims to distribute a fixed sum across diverse assets, aiming to maximize the portfolio's expected return while prudently managing risk.
Although classical methods such as the Markowitz model \citep{Markowitz} have been proven to be effective in the past, they rely on a static approach that may not be suitable for a rapidly changing market. 
Hence, introducing a dynamic portfolio optimization algorithm capable of adapting to changing market conditions is crucial for better risk management and return improvement.
We review bandit algorithms in Section \ref{sec:bandits_PM} and deep RL algorithms in Section \ref{multi-stage-pm}.


\subsubsection{Bandit Algorithms for Portfolio Management}\label{sec:bandits_PM}
The bandit problem \citep{lattimore2020bandit} is a special case of RL, in that there does not exist a state transition and, hence, there is no long-term dependency. 
Bandits algorithms have been studied in a few papers on portfolio optimization application \citep{shen2015portfolio, shen2016portfolio, zhu2019adaptive, huo2017risk}, where the investor dynamically updates the portfolio based on historical observations.


\cite{shen2015portfolio, shen2016portfolio, zhu2019adaptive} all adopt a standard bandits algorithm named Thompson Sampling ({TS}) \citep{russo2018tutorial} in a non-contextual multi-armed bandit setting, with the major difference in how to define the arms. 
\cite{shen2015portfolio} is the first work on bandits for portfolio optimization. 
This work focuses on the challenge that it is not appropriate to regard stocks as independent arms (due to the correlation between them). 
Towards risk diversification, the authors propose to integrate bandits with a popular practice in portfolio optimization, that is, first applying principal component analysis to the stock returns and then regard the eigenvectors as arms. 
Therefore, the paper addresses the limitation of a naive application of bandit algorithms in portfolio optimization and provides a nice integration with finance theory, which is further supported by the performance of the backtest when compared to several other online portfolio optimization algorithms. 
\cite{shen2016portfolio} similarly apply TS to portfolio optimization. 
The article studies portfolio blending, i.e., the authors regard the portfolios given by different selection strategies as arms and aim to find the best combinations of these strategies. 
The algorithm in \citet{shen2016portfolio} is designed to mix two base strategies, which is further extended in \citet{zhu2019adaptive} to allow mixing multiple base strategies. 
Superior performance over several online portfolio optimization algorithms as well as the base strategies is reported in backtesting.

The aforementioned works focus on maximizing the cumulative expected return while ignoring another important metric in finance, risk. 
To fill this gap, \citet{huo2017risk} propose to consider a linear combination between a stock selected using a bandit algorithm and another portfolio solved by (greedy) online portfolio optimization. 
The former targets the expected return (in the long run) and the latter is selected by minimizing the risk. 
By choosing the weights of the two parts carefully and manually, the paper aims to achieve a balance between the expected return and the risk. 









In closing, we raise a fundamental question: \textit{ is the bandit problem ideal for portfolio optimization, in other words, is active exploration necessary in this application?}
As highlighted by \cite{russo2018tutorial} (see Section 8.2.1 therein), this question warrants further discussions, which we delve into in Section \ref{sec: challenge}. 




\subsubsection{RL Algorithms for Portfolio Management}\label{multi-stage-pm}

In contrast to bandit algorithms, which focus solely on immediate rewards, RL algorithms approach the sequential portfolio management problem by formulating it as a Markov Decision Process (MDP).
By doing so, these algorithms take into account the consequences of current portfolio decisions on future states, making them more comprehensive and forward-looking. 
This sub-section will review papers that use the RL algorithms for training trading agents in portfolio management.

To begin with, \citet{jiang2017deep} propose Ensemble of Identical Independent Evaluators ({EIIE}), which is an ensemble of neural networks with history price as input, to solve the asset allocation problems in the cryptocurrency market.
The study explores three backbone networks (RNN, LSTM, and CNN) as RL policy networks and demonstrates that RL models can outperform traditional approaches, indicating their potential for portfolio management. Subsequent studies in RL algorithm trading agents consider this paper as a benchmark.

\paragraph{Complicated Policy Networks} 
\cite{liu2018practical} builds upon \cite{jiang2017deep}, by adopting DDPG over PG algorithms for trading agents optimization,  resulting in remarkable performance improvement.
This paper sheds light on how deep policy networks can improve the RL agents for PM tasks. 
More complicated policy networks (such as DQN, hierarchical agents, and parallel agents) are explored in  other studies.
For example, \citet{gao2020application} utilizes DQN for portfolio management with discrete action space, where
Duel Q-Net (Wang et al. 2015) is used as the backbone framework of the DQN. To improve sampling efficiency, this paper also employs Prioritized Experience Replay with the SumTree structure \citep{Schaul2015PrioritizedER}  to prioritize valuable experiences over noise during training.
\citet{wang2021commission} propose a Hierarchical Reinforced Portfolio Management (HRPM) approach.
Specifically, the hierarchical structure includes two agents: a high-level RL agent maximizing long-term profits, and a low-level RL agent minimizing trading costs while achieving the wealth redistribution targets set by the high-level model within a limited time window. 
Similarly, \citet{ma2021parallel} propose an approach with two parallel agents: one capturing current asset information and the other using LSTM layers to detect long-term market trends. 
More papers \citep{wang2019alphastock, lee2020maps} adopt multi-agents policy networks, which we will discuss further.

\paragraph{Empirical financial strategy oriented methods} 
Some studies have recognized the value of leveraging pre-existing domain knowledge to enhance the efficiency and effectiveness of RL models. 
Traditional financial strategies, like \textit{buying winners and selling losers} (BWSL) \citep{Jegadeesh1993ReturnsTB}, \textit{Rescorla-Wanger model} \citep{rescorla1972theory}, \textit{Modern Portfolio Theory} (MPT)\citep{menezes1970theory, pratt1978risk} and \textit{Dual Thrust} strategy are employed in work below.

AlphaStock \citep{wang2019alphastock} adopts the BWSL strategy, buying assets with high price rising rates and selling those with low rate. The paper designs two optimizing agents: a long agent purchasing selected winner assets and a short agent selling selected loser assets, optimizing for the Sharpe ratio 
using a cross-assets attention network.
Sensitivity analysis on stock feature shows that AlphaStock tends to select stocks with high long-term growth, low volatility, high intrinsic value, and being undervalued.
\citet{li2019optimistic} notes that a bear market, marked by pessimism, aligns with economic downturns, while a bull market, marked by optimism, occurs when security prices outpace interest rates. 
This paper incorporates market sentiment by modifying the Rescorla-Wanger model to adapt differently under positive and negative environments.
Modern Portfolio Theory (MPT)\citep{menezes1970theory, pratt1978risk} is widely used to construct optimal investment portfolios by balancing the trade-off between risk and return. 
\citet{zhang2020deep} construct the reward function of RL framework by leveraging MPT, where it shows that the constructed reward function is equivalent to the MPT's utility function subject to a proposed condition. 
Another strategy, Michael Chalek's Dual, is an investment approach combining the true range (a measure of volatility) with the dual thrust model, which sets Buy and Sell Thresholds. Specifically, a buy signal is generated when the market price exceeds the Buy Threshold, and a sell signal is triggered when it falls below the Sell Threshold.
\citet{liu2020adaptive}  utilize the Dual Thrust strategy  to initiate the demonstration buffer, enhancing  the deterministic policy gradient method to balance exploration and exploitation effectively.

\paragraph{Methods aimed at enhancing robustness}
The volatility of the financial market presents significant challenges when applying RL to real-world investment processes. 
To ensure consistent performance, some studies prioritize enhancing robustness, i.e. the stability of the model's performance under different market conditions, as a key objective,.
More discussion on the challenges of robustness can be found in Sec \ref{sec5:rob}.

\citet{yang2020deep} propose an ensemble strategy that can dynamically select from three RL algorithms (PPO, A2C, and DDPG) based on market indicators.
Experimental results in their paper show that this approach effectively preserves robustness under different market conditions.
\citet{lee2020maps} also focus on dealing with the continuously changing market conditions. 
Multiple agents are designed, each agent aims to optimize their portfolio while considering the potential negative impact on the overall system's risk-adjusted return if portfolios are similar.
Another robustness-related RL method is DeepPocket \citep{soleymani2021deep}. 
In DeepPocket, the interaction between assets is represented by a graph with financial assets as nodes, and the correlation function between assets as edges. 
The graph is encoded as the state in the actor-critic RL algorithm. The paper shows that the DeepPocket can handle unanticipated changes generated by exogenous factors such as COVID-19 in online learning.
\citet{benhamou2021detecting} suggests a multi-network approach that combines financial contextual information with asset states to predict the economy's health, which enables the model to outperform baseline methods expecially during recessions.


\paragraph{Market prediction based methods}
An emerging trend observed among many papers is to include predicted contextual information, such as \textit{asset price movements}, \textit{data augmentation}, \textit{market trend}, \textit{market sentiment}, or \textit{economic conditions}.

Market-prediction-based deep RL in the portfolio management domain was first applied in \citet{yu2019model}. This paper implements a prediction module to forecast the next time-step price and a market indicator aims to augment state space. Besides, the authors also propose 
a behavior cloning module to reduce volatility by preventing a large change in portfolio weights based on imitation learning.  Similarly, the data augmentation idea is explored in \citet{theate2021application} with DQN policy network.
Other than prices, more complicated state augmentation haven been explored.  
\citet{lei2020time} enhances the RL states by incorporating a prediction-based auto-encoder using the attention mechanism, to capture the market trend. 
\citet{ye2020reinforcement} incorporate the market sentiment feature extracted from relevant news as an external stock movement indicator within the RL framework. 
\citet{koratamaddi2021market} also take market sentiment into account, by augmenting the state space with a calculated market confidence score derived from company-related news and tweets.
DeepTrader \citep{wang2021deeptrader} is another market-prediction-oriented work. 
The authors design an asset scoring unit to capture asset-specific growth and a market scoring unit to adjust the long/short proportion based on the overall economic situation, which enables the model to balance risk and return and to adapt to changing market conditions.

%% file: 3_4_OE.tex
\subsection{Applications in Optimal Execution}\label{sec:OE}
Optimal execution ({OE}) is another important problem in finance \citep{bertsimas1998optimal}. 
Compared with portfolio optimization, OE is finer-grained: it focuses on the optimal way to buy or sell some pre-specified units of a single stock in a given time frame. 
Therefore, OE becomes an indispensable component of the continuous updating of portfolios. 
Since the naive strategy that places the entire order instantly may have a significant impact on the market and hence profitability, a well-designed algorithm is required. 

In the literature, 
various metrics are considered to evaluate OE algorithms, including the PnL \cite{nevmyvaka2006reinforcement, deng2016deep, wei2019model, shen2014risk}, a normalized version of the PnL called the \textit{implementation shortfall} \citep{lin2020E2E, hendricks2014reinforcement}, the Sharpe ratio \citep{deng2016deep} and the shaped reward \citep{fang2021universal, lin2020deep}. 


\textbf{Model-free RL. }
\citet{nevmyvaka2006reinforcement} is the first attempt to apply model-free RL to OE. 
Using a modified Q-learning algorithm, the paper conducts  numerical experiments on large-scale NASDAQ market microstructure datasets and reports significant improvements over baselines on trading costs. 
\citet{lin2020deep} similarly applies a variant of DQN to OE. 
The main novelty is to propose a shaped reward structure and incorporate the zero-end inventory constraint into DQN. 
Backtesting shows the DQN-based approach outperforms a few baselines, and the constraint improves the stability. 
In \cite{lin2020E2E}, the authors further design an end-to-end framework utilizing LSTM to automatically extract features from the time-dependent limit-order book data, demonstrating improvements over previous methods. 
Similar ideas are studied in \cite{deng2016deep}, by integrating RNN and policy gradient. 
In financial problems, The data is typically noisy with imperfect market information. 
To overcome this bottleneck, \citet{fang2021universal} proposes an oracle policy distillation approach: 
during training, a teacher policy is learned with access to the future price and volume, which is then distilled to learn the student policy, which has no access to future information and is the one to be deployed. 
Such a procedure reduces the training variance and leads to a better policy in testing, as demonstrated via experiments over a few  baselines. 

As discussed in previous sections, one prominent limitation of standard RL formulations is that the objective ignores the risk, which is of particular importance in OE and is also one of the main concerns of algorithmic trading. 
To fill the gap, 
\citet{shen2014risk} adopt the risk-sensitive MDP formulation in \cite{shen2013risk} to design a Q-learning-type algorithm. 
Specifically, the expected cumulative reward is replaced by a utility-based shortfall metric proposed in the mathematical finance literature. 
The backtesting results demonstrate improved robustness, especially during the 2010 flash crash.


\textbf{Model-based RL. }
The papers above rely on model-free RL and may suffer from low sample efficiency. 
\citet{wei2019model} firstly apply model-based DRL to OE: 
it first learns an environment model (simulator) using RNN and auto-encoder, and then trains a policy by running model-free DRL (including Double DQN, PG, and A2C) within the simulator. 
One main contribution of the paper is to show the algorithm's five-day real performance in the real market, where the agent is profitable and the model-predicted trajectory is close to the observed trajectory. 
In contrast, 
\citet{hambly2021policy} consider a structural model assumption, the linear–quadratic regulator problem. 
This paper uses the OE problem as an example to demonstrate the robustness of running model-free methods in a model-based environment over directly solving the model-based controller, when there is model misspecification. 

The model fidelity is typically central to the performance of model-based RL. 
\citet{karpe2020multi} utilize a flexible agent-based market simulator, ABIDES \citep{byrd2019abides}, to configure a multi-agent environment for training RL execution agents and show it converges to a Time-Weighted Average Price strategy. 

\textbf{Leveraging mathematical finance models. } 
The RL methods surveyed above are largely disconnected from the mathematical finance literature and therefore may not utilize some domain-specific structure for more efficient learning. 
Meanwhile, \citet{hendricks2014reinforcement} focus on optimal  liquidation and leverage the Almgren-Chriss solution \citep{almgren2001optimal} as a base algorithm: 
at every decision point, authors first compute the recommended action based on the Almgren-Chriss model, 
then allow the RL agent to learn an action as a multiplicative factor on this base action based on the market microstructure state vector. 
With the additional flexibility, the RL agent successfully improves the shortfall by 10.3\% compared with the Almgren–Chriss solution. 
One limitation of this analysis is the missing of comparisons with baseline RL agents. 




\textbf{Inverse RL. }
The reward definition quality is also critical to the empirical performance of RL agent. 
\citet{roa2019towards} study revealing the reward function of expert agents by inverse RL. 
This work applies three inverse RL algorithms to simulated datasets to study their performance, and identifies the failure of linear model-based inverse RL algorithms in recovering non-linear reward functions. 

%% file: 4_Meta_Analysis.tex
\section{Meta Analysis of Experimental Results}\label{sec:meta-analysis}

In this section, we conduct several meta-analyses on the effectiveness of RL in financial domains. 
Our objective is to explore four key research questions, which we believe can help analyze the effectiveness of RL methods in finance in current literature, discover challenges in RL application (Section \ref{sec: challenge}),
and provide insights for future research directions (Section \ref{sec: future}).
(1) the relationship between MDP design and model performance, 
(2) the influence of training duration on model performance, 
(3) the most prevalent RL algorithms and their relative performance, and
(4) the suitability of current studies to the financial domain.
To answer these questions, we collect a comprehensive set of literature  summarized in Table \ref{table:state_action}
and analyze the experimental results reported in these papers.

\begin{figure*}
\centering
\subcaptionbox{Linear regression to analyze the relationship between RL Premium and feature dimensions, where the slope reveals the correlation strength and the p-value signifies the significance of this linear connection. \label{pm-feature}}[.45\linewidth]{
    \includegraphics[width=\linewidth]{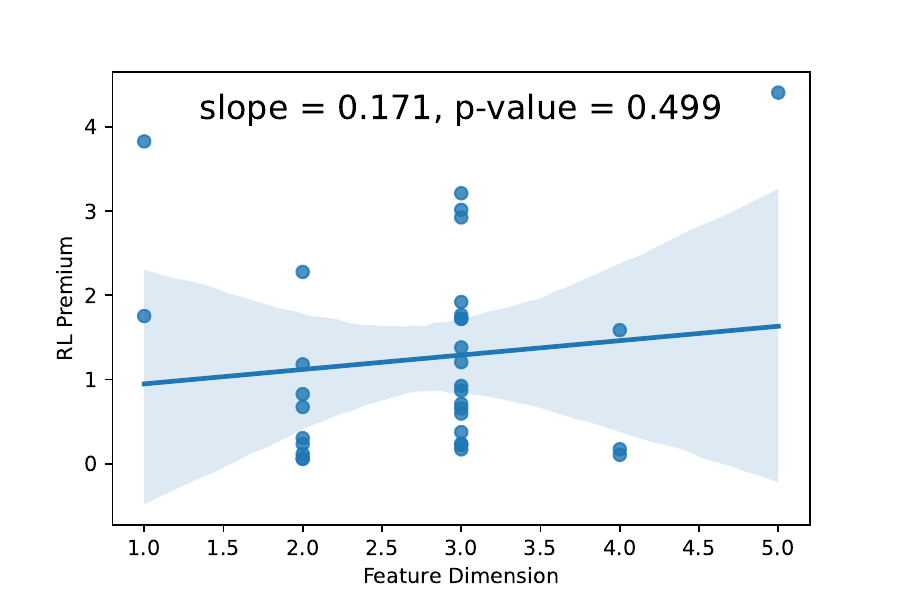}
\vspace{-15pt}
} 
\hfill
\subcaptionbox{Linear regression to analyze the relationship between RL Premium and number of assets. \label{pm-assets-count}}[.45\linewidth]{
    \includegraphics[width=\linewidth]{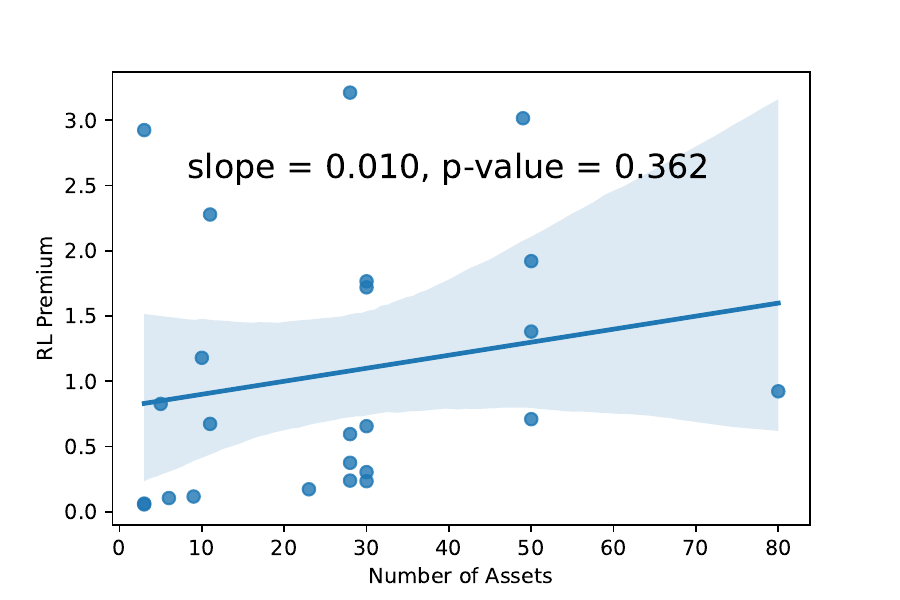}
\vspace{-15pt}
}
\subcaptionbox{Box plots comparing RL premium
based on whether the reward model use return or shaped return, with significance of
difference tested by a two-sample t-test
assuming same variance. \label{pm-reward-performance}}[.45\linewidth]{
    \includegraphics[width=\linewidth]{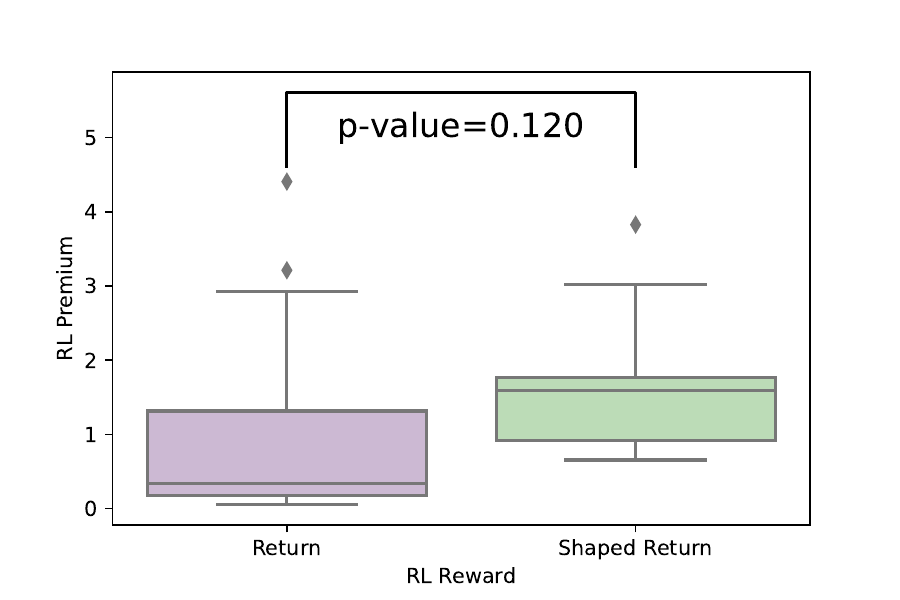}
\vspace{-15pt}
}\hfill
 \subcaptionbox{Linear regression to analyze the relationship between RL Premium and years of training period, where the slope reveals the correlation strength and the p-value signifies the significance of this linear connection. \label{pm-training}}[.45\linewidth]{
        \includegraphics[width=\linewidth]{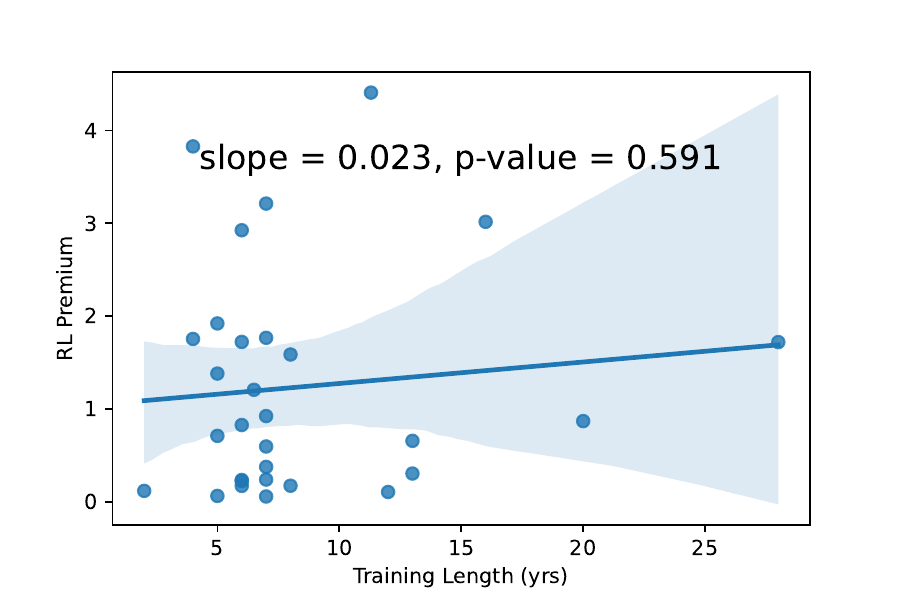}
\vspace{-15pt}
    }
\subcaptionbox{Box plots comparing RL premium based on whether the training period includes a recession, with significance of difference tested by a two-sample t-test assuming same variance. \label{pm-recession-performance}}[.45\linewidth]{
        \includegraphics[width=\linewidth]{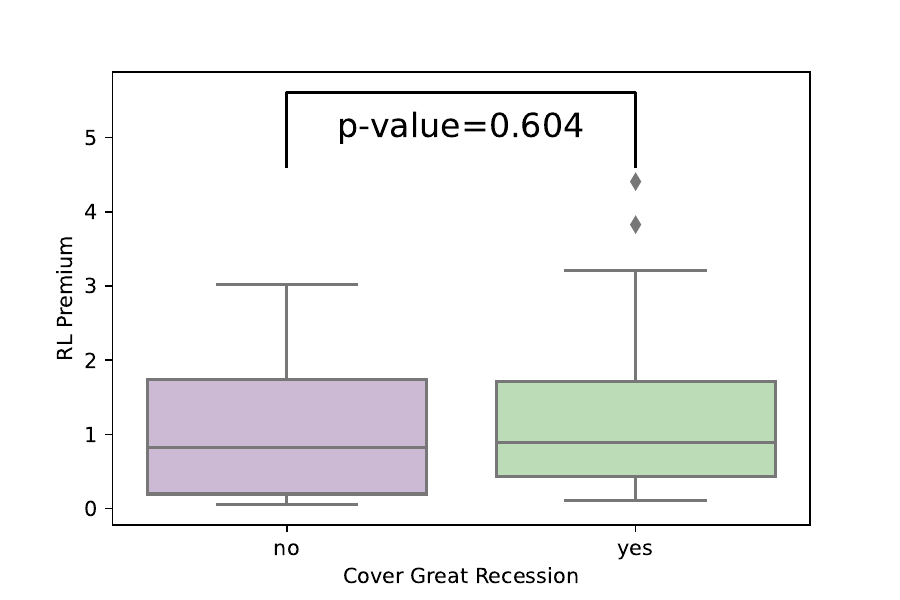}
\vspace{-15pt}
    }    \hfill
\subcaptionbox{Box plots analysis of RL premium, contrasting outcomes from RL PG and DQN strategies, with significance tested by a two-sample t-test.\label{pm-algo-performance}}[.45\linewidth]{
        \includegraphics[width=\linewidth]{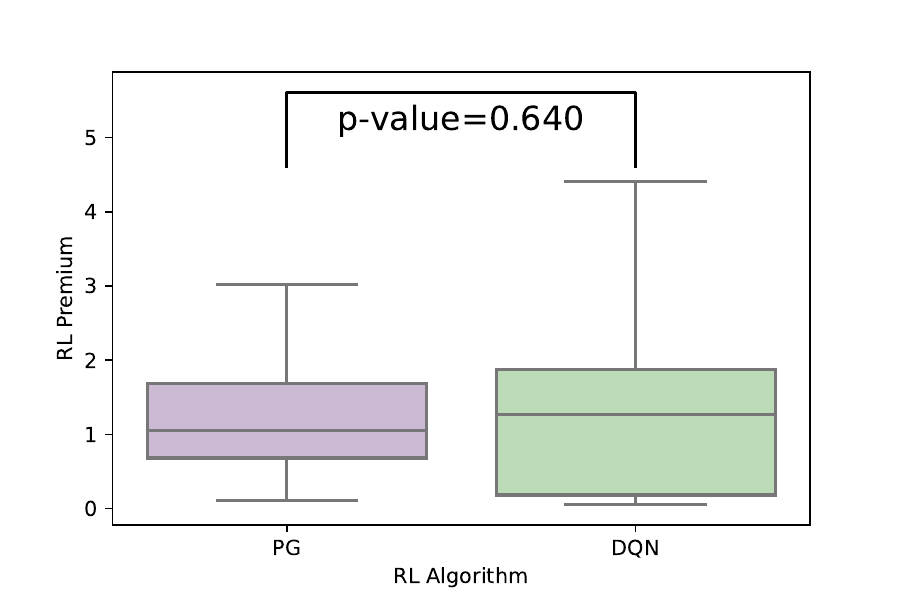}
\vspace{-15pt}
    }
\\[10pt]
\caption{RL Premium Analysis.}
\vspace{-10pt}
\end{figure*}




The varying use of test data and baselines in different research papers presents a significant challenge in conducting a meta-analysis. 
To address this, we proposed the \textit{RL Premium} metric to benchmark model performance consistently across studies, providing insights into performance-influencing factors and guiding future research. Although it has broad applications across the financial field, it is particularly useful in portfolio management, where the Sharpe ratio is a standard evaluation metric. 
While alternative metrics like PnL exist for comparing model performance, the proposed \textit{RL Premium} metric offers distinct advantages in ensuring fair comparisons.
By mitigating variations in time periods, markets, and transaction cost across different studies, our metric provides a more standardized basis for evaluation.
The \textit{RL Premium} is calculated by subtracting the Sharpe ratio of the strongest non-RL baseline $\text{SR}_{\text{Baseline}}$ from the RL method's Sharpe ratio $\text{SR}_{\text{RL}}$, then normalizing this difference by $\text{SR}_{\text{Baseline}}$.
Formally, the RL premium observed in a paper is defined as\footnote{In portfolio management analysis, RL premium that falls outside of the range between $5\%$ quantile and $95\%$ quantile of the whole sample are treated as outliers and hence removed.}  
$$\text{RL  Premium} = (\text{SR}_{\text{RL}} - \text{SR}_{\text{Baseline}}) / \text{SR}_{\text{Baseline}}. $$ 
For example, in \citet{jiang2017deep}, the strongest non-RL baseline has a Sharpe ratio of 0.012, while the RL method has a Sharpe ratio of 0.087. 
The RL premium in this case would be calculated as $(0.087 - 0.012) / 0.012 = 6.25$, indicating a significant performance gain from the use of the RL method.

  




\textbf{Q1: Does MDP design affect RL performance?}
We examine the relationship between RL Premium and the design of MDPs, specifically the state, action, and reward components. 

\textbf{State}: 
We observe a growing trend of integrating supplementary information beyond asset prices into the state space.
For instance, \citet{zhang2020deep} includes technical indicators as part of the state for the RL agent and \citet{koratamaddi2021market} considers market sentiment. 
To study whether including more features can help with model performance, we regress the RL premium on state feature dimension in Figure \ref{pm-feature}, with each experiment setting treated as separate data point. 
Findings suggest that most papers use two to three features, with only a few incorporating more. While upward slope demonstrates that additional information may slightly improve portfolio management performance, the effect is not significant based on p-value.

\textbf{Action}: 
In multi-asset PM tasks, the number of assets is related with the size of action space.
As illustrated in Figure \ref{pm-assets-count}, RL models perform well when the action space is extensive.
While optimizing numerous assets manually by human analysts can be a daunting task, RL models face a less arduous challenge in this regard\footnote{To prevent skewness from affecting the analysis, studies with over 100 assets were excluded.}.

\textbf{Reward}: Designing a suitable reward is essential in RL modeling as it can greatly uplift decision-making efficiency.
In portfolio management, the most commonly used reward is the portfolio return.
Innovative rewards such as risk-adjusted or sentiment-adjusted returns have been proposed in some studies.
To understand the impact of rewards, we compare the RL Premium using the original return and the shaped return. 
Figure \ref{pm-reward-performance} shows that the use of shaped rewards leads to a noticeable improvement in the performance of the model.

\textbf{Q2: How does the training period affect model performance?}
The volatility of financial market data highlights the importance of carefully choosing the training period.
Our analysis reveals that the length of the training period varies by study, indicating the absence of established standard. 
We are interested to understand whether longer training period will improve or harm model performance.
However, from regression shown in Figure \ref{pm-training} we cannot conclude the relationship between the two. 
One possible explanation is that financial information is rapidly changing, so a lengthy training period may introduce noise rather than offer useful information for recent decisions.

Including a recession period in training data poses challenges for RL models, as recession can lead to sudden and unpredictable market changes, rendering non-stationary conditions and hindering precise decision-making.
However, statistical analysis leveraging a two-sample t-test to evaluate differences in RL premium for periods covering a recession in Figure \ref{pm-recession-performance} shows that inclusion of a recession period does not appear to negatively affect the RL Premium.

\textbf{Q3: Does the choice of RL algorithm have any impact on model performance?}
Among all the articles we have reviewed, PG and DQN are the most popular RL algorithms in portfolio management, as shown in Table \ref{table:reward+algo} in the Appendix \ref{sec:appendix_ref_table}. To compare the difference, we perform a two-sample t-test on the average RL premium between the RL methods PG and DQN.
From Figure \ref{pm-algo-performance}, no significant differences are detected between the two algorithms. In the literature on market making, the RL premium of three critic-only methods (Q-learning, SARSA, and R-learning) can be calculated.

\begin{figure*}[hb]
    \centering
    \subcaptionbox{Distribution of papers based on the assumption of zero versus non-zero slippage. \label{pm-slippage}}[.45\linewidth]{
        \includegraphics[width=\linewidth]{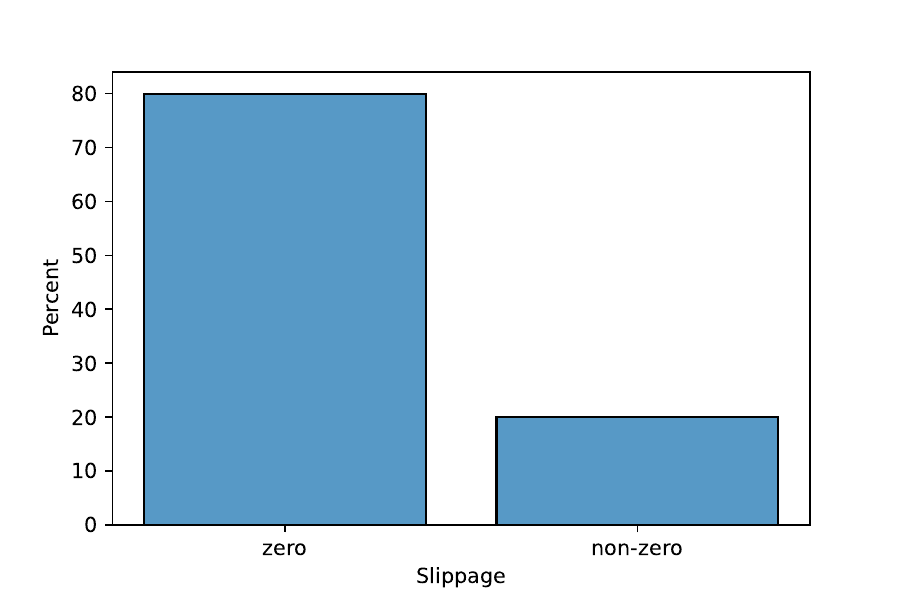}
    \vspace{-15pt}
    }\hfill 
    \subcaptionbox{Distribution of papers categorized by transaction costs assumptions.\label{pm-commission}}[.45\linewidth]{
        \includegraphics[width=\linewidth]{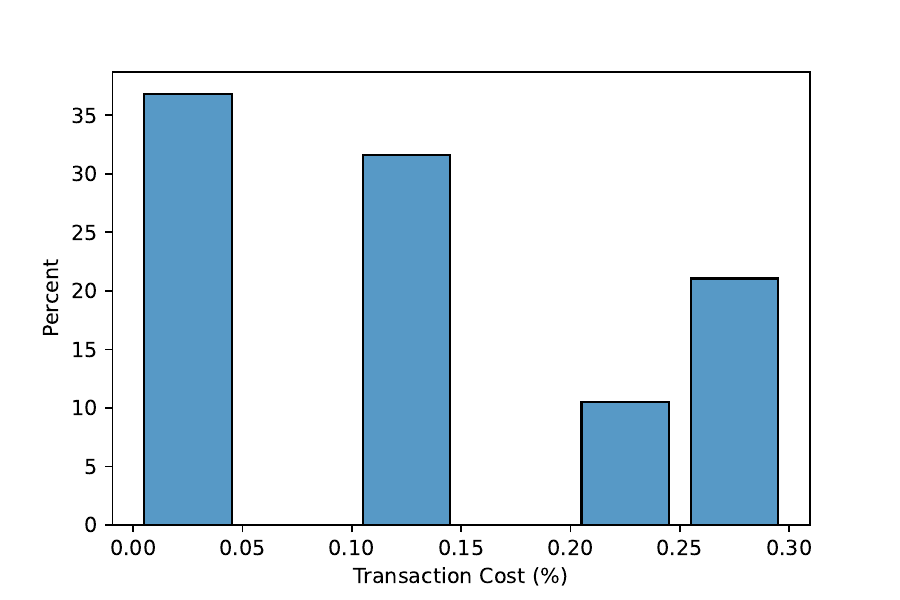}
    \vspace{-15pt}
    }
    \\[10pt]
    \caption{Analysis of realistic assumptions in surveyed papers. }
    \vspace{-10pt}
\end{figure*}

\textbf{Q4: Do the assumptions made in the literature accurately reflect reality?}
In practical portfolio management, certain constraints, such as slippage and transaction costs, can negatively impact investment returns. Neglecting these real-world constraints may lead to an unrealistic evaluation of the performance of the RL model. 
We notice that most papers either assume zero slippage or do not discuss the treatment in their models (as shown in Figure \ref{pm-slippage}). Regarding the transaction cost, the most common range used in the literature is 0.2-0.3$\%$ (as shown in Figure \ref{pm-commission}). 

%% file: 6_env.tex
\section{Discussions on Environments, Packages and Benchmarking}\label{sec: data discussion}
High-quality datasets/environments, rich open-sourced packages, and fair benchmarking are critical for the advance of an RL area. 
In subsection \ref{sec:env}, we give a detailed discussion of the training and evaluation environments used in RL for finance.
Subsection \ref{sec:Benchmarking} reviews major open-sourced packages and platforms, and discusses the benchmarking status. 

\subsection{Discussion on Training and Evaluation Environments}\label{sec:env}
Interactive data collection is crucial for training or evaluating RL agents. This section covers various environments and their uses. 
Among all the papers we survey, the environments fall into three categories: 
(i) the real financial market,  
(ii) a historical trajectories-based simulator with minimal assumptions,  
and (iii) a simulator with a manually designed data generation mechanism. 
While real market data carries high requirements, we primarily focus on the latter two types. Nevertheless, incorporating more real-world evidence could significantly advance the field.

\subsubsection{Historical trajectories-based simulator and exogenous MDPs}\label{sec:exogenous}
This section examines the assumptions behind replaying historical trajectories in RL and its technical validity. The exploration-exploitation trade-off, a core RL challenge, arises because actions not taken provide no outcomes (i.e. counterfactuals), making direct simulation from historical data without assumptions impossible. 
Fortunately, in MDPs with exogenous inputs (\textit{Exo-MDP}, \citet{chitnis2020learning}), 
simulating by re-playing historical trajectories turns out feasible. 
Specifically, suppose that the state variables can be decomposed into two parts as $s_t = (x_t,z_t)$, where
(i) $P(x_{t+1}|s_t, a_t)= P(x_{t+1}|x_t)$ and (ii) $z_{t+1} = f(s_t,a_t)$ with $f$ as a known function. 
Then, at time $t$, even if we take an action $a'_t$ in state $s'_t$ and both are different from the recorded ones, setting $s'_{t+1} = (x_{t+1}, z'_{t+1})$ with $z'_{t+1} = f(x_t, z'_t, a'_t)$ is a valid way to sample the next state. 

In the literature we reviewed, state variables in finance are often grouped into external information (like news or the day of the week), public market data (such as price or volume), and private agent states (like inventory levels). The market data is typically considered exogenous. The effect of an agent's actions on external information is termed the "market impact." 
In many finance applications, Exo-MDP is a reasonable assumption, hence providing the validity of the widely used backtesting-type protocol. 
For example, most portfolio optimization papers surveyed in Section \ref{sec:Portfolio Management} presume that only stock prices and side information are random, the agent's actions minimally affect the market, and state components related to the agent's own portfolio can be updated by accounting in a predetermined and deterministic manner. 
Therefore, the Exo-MDP assumptions are satisfied, and we can use the historical trajectories to evaluate any policy, for either learning or evaluation purposes. 
There is minimal bias since this method does not require additional model assumptions like those needed in designed simulators.


However, Exo-MDP may not be a suitable model in some other applications such as optimal execution (see Section \ref{sec:OE}), where the market impact is a central topic. 
Interestingly, many papers on OE often overlook the market impact when using historical data for training or backtesting. 
We recommend that a more thorough discussion and justification of using historical data in contexts like optimal execution is a critical next step. Additionally, developing a high-fidelity simulator, similar to ABIDES \citep{byrd2019abides}, would be highly beneficial for these studies.


To conclude, we refocus on the bandit algorithms for PM  discussed earlier in Section \ref{sec:bandits_PM}. In the surveyed literature, experiments utilize historical trajectories assuming an exogenous price process. Consequently, active exploration is unnecessary, suggesting that traditional online portfolio selection \citep{li2014online} algorithms might suffice. However, the rationale for employing bandit algorithms in this context is not well articulated, and further in-depth discussions could clarify their relevance and utility.

\subsubsection{Simulator with a manually designed data generation mechanism}\label{sec:simulator}
In applications where the assumption in Section \ref{sec:exogenous} cannot be satisfied or when high-quality data is not available, 
a simulator with a manual-designed data generation mechanism is typically used. 
As widely acknowledged in the RL field, the simulator quality is often critical to the learning performance: 
a high-fidelity simulator can make the exploration much easier by generating more data when needed, 
while a simulator far from the real environment may cause a significant sim-to-real gap \citep{zhao2020sim}, i.e., a learned policy that performs well in the simulation may fail once deployed in the real world. 


In the finance domain, we observe two major types of data generation model: 
\textit{probabilistic model-based} and \textit{agent-based}. 
The former is developed on the basis of the rich mathematical finance literature \citep{horst2018introduction}, e.g. by modeling the financial time series by some stochastic processes or other models from partial differential equations; 
the latter focuses on the fact that the finance market is composed of many trading agents who jointly determine the dynamic, and hence this approach chooses to model the trading behavior of these agents \citep{byrd2019abides}. Both approaches are popular and sometimes even used together. 


Despite the importance discussed above, high-fidelity simulators have not been publicly available in previous years, possibly due to the sensitivity and privacy nature of the finance domain. 
The lack of a fair evaluation and comparison benchmark largely impedes the development of the area. 
Fortunately, in recent years, there has been increasing attention to this issue, with several benchmark environments published \citep{byrd2019abides, liu2021finrl, ardon2021towards}. 
We do believe that these advances would be as valuable and critical as those in methodology innovation.

\subsection{Open-source packages and Benchmarking}\label{sec:Benchmarking}
In Figure \ref{fig:method}, we summarize the frequency of different RL algorithms used in finance. 
Although there are works such as \citet{Spooner2018MM} comparing different RL algorithms under the same settings, 
there are still no comprehensive comparisons of common model-free algorithms (actor-only, critic-only and actor-critic). 
Among the papers reviewed, different works adapt different MDP processes (either generated by simulator or from real data) and RL approaches. Thus, to obtain a benchmark of different RL algorithms, a unified simulator is required. 
In the FinRL tool developed by \citet{liu2021finrl}, many different RL algorithms are implemented in the same environment for portfolio management, and such benchmarking can be achieved. However, in other areas such as market making, different tasks still adapt different price-generating processes and metrics. Specifically, most market-making papers report their PnL, but the scale of that is not unified, as discussed in the definition of ``RL premium''. Thus, a unified benchmarking study is needed to compare different RL algorithms proposed by different papers.

To implement popular RL algorithms, open-source packages such as RLlib \citep{liang2018rllib}, Stable-baselines \citep{raffin2021stable}, Tensorflow agents \citep{hafner2017tensorflow} can be used. However, to implement and adapt them to the finance domain, an end-to-end pipeline and platform is the key, but could be time-consuming and a major factor contributing to the inconsistency in settings between research papers. 
Several open-source packages have been contributed to this purpose, such as FinRL \citep{liu2021finrl2,liu2020finrl,liu2021finrl} and TensorTrader \citep{tensortrade}.

%% file: 5_challenges.tex
\section{Challenges} \label{sec: challenge}
Despite recent advances in RL methods in many finance areas \citep{hambly2021recent}, there are still many challenges to designing a practical RL system in the real world. 
Some of these challenges come from the nature of financial data, and others are due to the limitations of RL algorithms. 
We discuss these two aspects in Sections \ref{sec:challenge_data} and \ref{sec:limitation_algorithm}, respectively.



\subsection{Challenges due to Peculiar Features of Financial Data}\label{sec:challenge_data}

In finance, RL tasks require special consideration due to the distinct characteristics of financial data, as highlighted by \citet{fan2017elements}: high volatility, heavy-tailed distribution, non-stationarity, and asymmetry.

\subsubsection{High volatility}\label{subsec:high_volatility}


In finance, ``high volatility" refers to significant fluctuations in asset prices. It's often measured by the standard deviation of an asset's price over time or by its "Beta" value, as described by \citet{ lakonishok1984stock}, which compares the stock's volatility to the overall market.

One approach to dealing with stock volatility in RL is to proceed with low and high-volatile stocks separately, as suggested by \citet{ jin2016portfolio}. The authors assume prior knowledge of which stocks are high and low volatility, and prevents the agent from trading both types simultaneously, potentially boosting Sharpe ratio and reducing variance. 
However, solely relying on known volatility levels may be inadequate, necessitating the prediction of future volatility. While machine learning offers methods for volatility forecasting \citep[see, e.g.][ for an overview]{ge2022neural}, integrating them into RL in finance remains largely unexplored.


\subsubsection{Heavy-tailed distribution}\label{subsec:heavy_tail}



The tail of the probability distribution of stock returns (defined as the ratio of price change over the initial price) is typically heavier than that of a normal distribution \citep{bradley2003financial}. For instance, \citet{fan2017elements} provide examples of some stock returns in S$\&$P 500 whose distribution tails are heavier than a t-distribution with degree $5$, indicating that events far from the distribution mean may occur more frequently than expected under a normal distribution. In the context of RL, this phenomenon corresponds to cases where the reward distribution has a heavy tail, as the reward is calculated based on stock returns.



Although the heavy-tail issue has long been observed in financial data, in the RL literature, researchers still typically assume a light-tailed distribution, such as the Gaussian distribution, for rewards.
To our knowledge, \citet{zhuang2021no} is the only work that has designed RL algorithms for the heavy-tailed reward distribution setting. They propose a Q-learning algorithm that can achieve near-optimal regret bounds in heavy-tailed settings. Moreover, they have shown that learning the reward distribution is even more challenging than learning the state transition in such a setting.
Since this paper is relatively recent, RL algorithms that are robust to heavy-tailed distributions have not yet been widely applied in finance applications.

\subsubsection{Non-stationarity}

The empirical validity of most state-of-the-art RL algorithms heavily depends on the stationarity of the environment, in other words, the assumption that the system transition and the reward function are stable \citep{li2022reinforcement}. 
However, the non-stationarity of the finance market has been widely acknowledged \citep{fan2017elements}. 
For example, the market dynamic during the financial recession in 2008 was found to be very different from the normal period \citep{guharay2013analysis}; therefore, including this period in training may affect the performance of the tests in normal periods, as discussed in our meta-analysis (see Section \ref{sec:meta-analysis}). 

Up to the best of our knowledge, no work on finance RL has studied the non-stationarity issue formally. 
The training period is typically picked based on experience or data availability, without much justification.  
We believe that careful handling of this issue, for example, by testing stationarity and detecting change points \citep{li2022reinforcement}, can further improve the performance of RL in finance. 
A stronger connection to the vast literature on non-stationarity in financial time series would also be valuable \citep{schmitt2013non}. 

\subsubsection{Long-range dependency and latent information} 
Another critical assumption of most RL algorithms is the Markovian assumption, or roughly speaking, that our state variable contains sufficient historical information to predict the state transition. 
Without such an assumption, the foundation of these RL algorithms is shaken, which can lead to deterioration of their performance \citep{shi2020does}. 
Validating or satisfying this condition is particularly challenging in finance due to the market's complex dynamics and the lack of access to critical but unobservable information.
Moreover, it is widely observed that the financial time series has long-range dependency \citep{fan2017elements}. 

Based on our observation, the selection of state variables in financial RL works is largely based on the researchers' personal experience alone, without much explanation and validation. 
One common practice is to include a long time range of past data points, with the motivation to satisfy the Markovian assumption to a certain extent. 
For example, \citet{lin2020E2E}, \citet{lin2020deep}, and \citet{deng2016deep} use LSTM or RNN to learn from past time points. 
However, including a high-dimensional vector will make learning harder due to the introduced noise. 
Among the cited works, \citet{fang2021universal} propose an oracle policy distillation approach and \citet{liu2020adaptive} adopt POMDP algorithms to partially remedy information imperfectness. 
We suggest that one valuable future direction is to investigate which variables should be included in the state space. 
The Markovian assumption testing procedure and the variable selection procedure developed for RL in \citep{shi2020does} would be very useful.

\subsection{Challenges due to Limitations of RL Algorithms}\label{sec:limitation_algorithm}

Another set of challenges arises from the inherent constraints of RL methods. These intrinsic limitations include the lack of interpretability, issues with Markov Decision Process (MDP) modeling, and the models' lack of robustness and generalizability.

\subsubsection{Explainability}

Obtaining insightful interpretations from DRL methods presents significant challenges.
As \citet{israel2020can} point out that asset managers, driven by fiduciary duty, prefer interpretable and transparent models to communicate risks to clients. 
Therefore, the choice between interpretability and predictive power becomes a business decision for them. Recently, interpretable RL methods \citep{10.1145/3616864} have gained traction, suggesting a promising avenue for research in finance.




\subsubsection{MDP Modeling} 
The challenge of modeling a finance application as an MDP comes from two major aspects, first, how to select the variables considered in the States $\mathcal{S}$; second, how to decide the reward function $\mathcal{R}$.
Potential variables in financial data may include inventory, executed orders, historical prices, high-frequency data, and even news from the real world (Table \ref{table:state_action}). 
It is not realistic to fit all factors into the state due to the curse of dimensionality \citep{bellman1966dynamic}. 
However, enough information should be included in the MDP to make sure that RL can accurately predict expected rewards or/and next states given an action. 
The reward functions vary in different financial applications.
Even in the same financial application, no universal reward function is used in different papers. One reason is that in real-world financial applications, many metrics should be considered including but not limited to profit, market impact, and regulator's requirements. Therefore, it is challenging to combine the different types of outcomes into a single reward function.


\subsubsection{Robustness}\label{sec5:rob}

Robustness is one of the main focus of modeling, which indicates how stable the model performance could be under different conditions. 
As discussed in \cite{israel2020can}, the difficulties of using advanced models in financial markets, particularly in portfolio management, are largely due to the small data size and low signal-to-noise ratio. 
However, very limited attention has been paid to robustness.
\citet{wang2021commission} demonstrate the robustness by assessing the model performance across various markets and time periods. In general, we could not find a well-structured framework to measure and improve robustness in existing financial RL works.

%% file: 6_Future_Directions.tex
\section{Future Directions}
\label{sec: future}
Application of RL in finance has been an active research area in recent years. 
In this section, we discuss some future directions in this area.

\textbf{Multi-agent RL.}
Most reviewed articles use a single RL agent in financial applications. However, \citet{spooner2020robust} and \citet{gavsperov2021market} suggest that incorporating adversarial agents can enhance the robustness of the RL agent. This adversarial setting has also been explored in robotics and card games \citep{pinto2017robust,heinrich2016deep}. \citet{Patel2018MM} introduced a multi-agent framework with hierarchical policies, which is also applicable in complex RL problems beyond finance,  such as navigation and locomotion \citep{nachum2018data}. Despite these insights, multi-agent RL papers are a small portion of the reviewed works, indicating a need for further exploration.


\textbf{Model-based RL.}
As is pointed out by \citet{wei2019model}, there is an inherent connection between the model-based RL method and electronic trading systems, since electronic trading systems usually consist of a part to simulate the market. Therefore, it is very natural to use model-based RL algorithms that learn a simulator for the environment.
However, as shown in Figure \ref{fig:method}, few articles on optimal execution use model-based RL. More research is needed to apply model-based RL algorithms across various finance fields.


\textbf{Offline RL.} 
All surveyed works use \textit{online} RL algorithms, which rely on interactions with an environment to generate new trajectories and improve the agent. In contrast, \textit{offline} RL \citep{levine2020offline} learns directly from historical trajectories, without the need for new interactions. Offline RL has gained attention recently due to its safety and practicality in applications where online interaction is risky or infeasible. Moreover, Section \ref{sec:exogenous} suggests that using historical data as a simulator for online RL may be incorrect in settings with market impacts; in contrast,  offline RL is designed to handle these counterfactual issues. Therefore, offline RL is a promising direction for further research.

\textbf{Risk-Sensitive RL.}
Risk in investment refers to the uncertainty of capital loss. In portfolio management, many reviewed papers incorporate the Sharpe ratio in their reward functions to account for the volatility of return. In market making, inventory penalties are added to the reward functions for reducing risks. In summary, most reviewed papers of RL in finance still use traditional RL algorithms and address risk through reward function design. However, some RL papers manage risk via design of RL algorithms themselves, such as the risk-averse Q-learning method by \citet{shen2014risk}. Future works could incorporate risk-sensitive RL algorithms such as \citet{mihatsch2002risk,shen2014risk}. 




\section*{Disclaimer}
It is important to note that the results of the quantitative meta-analysis presented in this paper are intended to provide some general insight into the field of RL in finance. 
However, the results obtained are limited by the availability of data and variations in the settings used in each study. 
Furthermore, measurement of the relationship between model performance and factors only captures the impact of each factor independently and does not take into account potential confounders that may have influenced the outcome. 
Although the findings may be helpful in understanding the current stage of this research topic, the conclusions drawn from this study should be interpreted with caution and should not be considered definitive or conclusive.



%% file: 7_appendix.tex
\section{Reference Table}\label{sec:appendix_ref_table}

\pagenumbering{roman}

\input{Tables/table_combined}

\begin{figure*}[hb]
    \centering
    \subcaptionbox{Market Making\label{mm-method}}[.32\linewidth]{
        \includegraphics[width=\linewidth]{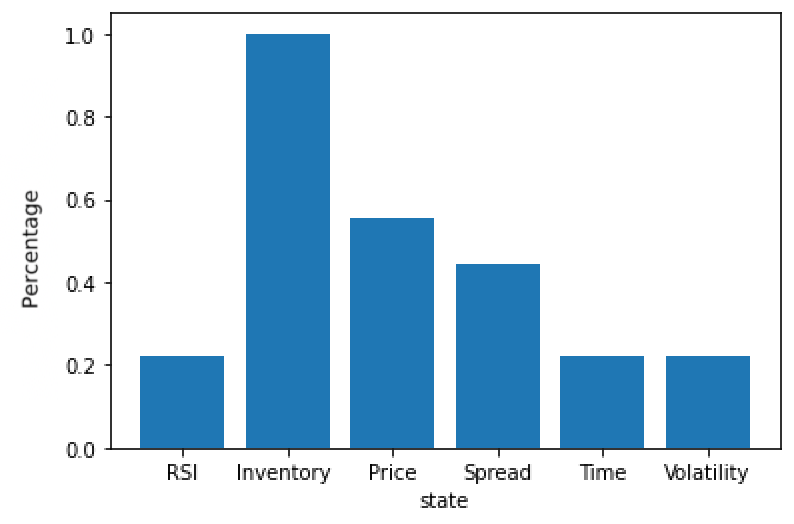}
    }\hfill 
    \subcaptionbox{Portfolio Management\label{pm-method}}[.32\linewidth]{
        \includegraphics[width=\linewidth]{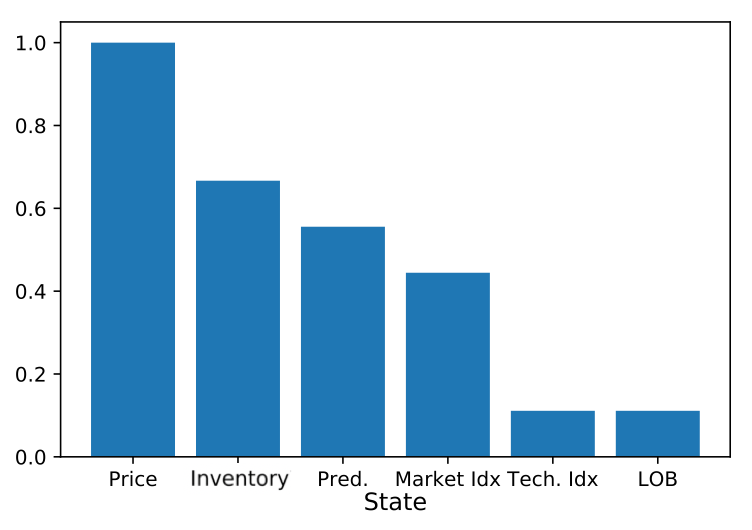}
    }\hfill 
    \subcaptionbox{Optimal Execution\label{fig:oe_state}}[.32\linewidth]{
        \includegraphics[width=\linewidth]{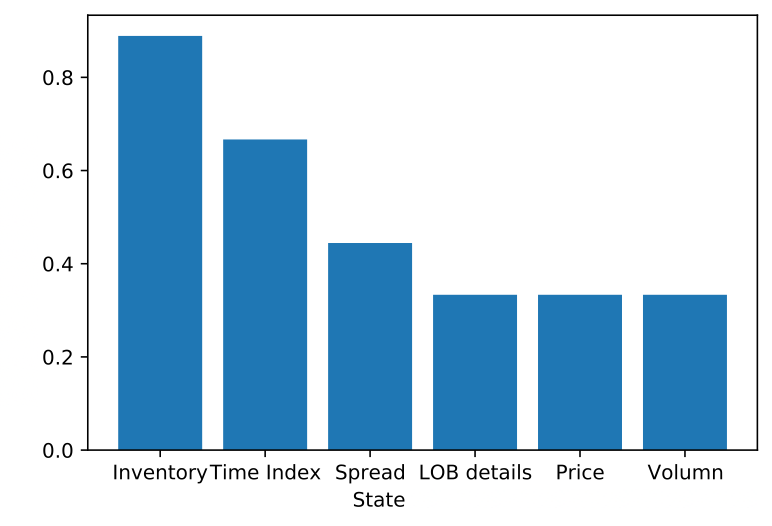}
    }   
    \\[10pt]
    \caption{Distribution of different state variables. Prices include the asset price, returns, and a combination; represents the agent's portfolio information; Pred. means the prediction results of stock movement and company news sentiment; Market Idx represents the market indicators like SP500; Tech. Idx means the technical indicators like Relative Strength Index (RSI); LOB means the limit-order-book is used as state; Time Index means the time interval is used as a state.}
    \vspace{-10pt}
\end{figure*}

\input{Tables/tab_simulator_MM}

\newpage
\clearpage

\section{Additional Meta-analysis Results for Market Making}\label{sec:appendix_more_res}

\begin{figure}[H]
\centering
\subcaptionbox{Linear regression to analyze the relationship between RL Premium and feature dimensions, in Market Making\label{mm-feature}}[.45\linewidth]{
    \includegraphics[width=\linewidth]{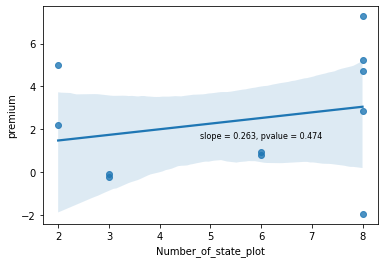}
}
 \subcaptionbox{Box plot analysis of RL premiu, of critic-only methods, in Market-Making.\label{mm-algo-performance}}[.45\linewidth]{
        \includegraphics[width=\linewidth]{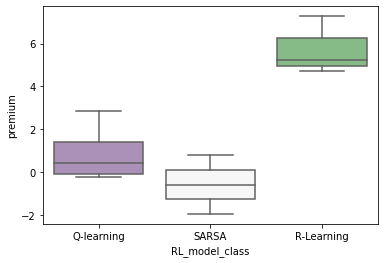}
    }
\\[10pt]
\caption{Impact of MDP Design on Model Performance.}
\vspace{-10pt}
\end{figure}

The RL premium on \textit{market making} can be defined in a similar way, where the Sharpe ratio is replaced by the profit and loss (PnL) of different methods, \begin{equation*}
    \text{RL\ Premium} = (\text{PnL}_{\text{Model}} - \text{PnL}_{\text{Baseline}}  ) / \text{PnL}_{\text{Baseline}}.
\end{equation*} 
In Figure \ref{mm-feature}, adding additional information to the model may slightly improve performance for portfolio management, although the impact is not substantial.
From Figure \ref{mm-algo-performance}, we observe that the performance of the R-learning methods is significantly better than those of the other two methods.  It may be interesting to explore more RL methods in future studies.

%% file: Tables/table_combined.tex

\begin{table}[H]
\caption{Literature on RL methods: state and actions. }
\vspace{.2cm}
\begin{tabular}{lp{0.4\textwidth}|llllllllll|llllll}
\hline
\multirow{2}{*}{}  & \multirow{2}{*}{Literature}  &  \multicolumn{10}{c|}{State} & \multicolumn{6}{c}{Action}\\ \cline{3-18} 
 &  & \rotatebox{90}{Price} & \rotatebox{90}{Inventory} & \rotatebox{90}{Market Indicator} & \rotatebox{90}{Tech-Indicator} & \rotatebox{90}{Company Info} & \rotatebox{90}{Predictions} & \rotatebox{90}{Bid-ask Info} & \rotatebox{90}{Trading Volume} & \rotatebox{90}{E2E} & \rotatebox{90}{Time} & \rotatebox{90}{Allocation Weight} & \rotatebox{90}{Long/Short} & \rotatebox{90}{Bid Price} & \rotatebox{90}{Quantity} & \rotatebox{90}{Place Limit Order} & \rotatebox{90}{Clear Inventory} \\ \hline
\multirow{20}{*}{\rotatebox{90}{Portfolio Management}} & \cite{jiang2017deep}  &  \cmark & \cmark  & & &  & & & & & & \cmark & & &  & &  \\
 &  \cite{yu2019model} &  \cmark & \cmark  & \cmark & &  & \cmark & & & & & \cmark & & &  & &  \\
 &  \cite{zhang2020deep}  &  \cmark & \cmark  & & \cmark  & & & & & & & & \cmark & &  & &  \\
 &  \cite{ye2020reinforcement}& \cmark & & & &  & & & & & & \cmark & & &  & &  \\
 &  \cite{wang2021deeptrader} & \cmark & & \cmark & &  & \cmark & & & & & \cmark & & &  & &  \\
 &  \cite{wang2021commission} & \cmark & \cmark  & & &  & \cmark & & \cmark & \cmark &  & \cmark & & &  & &  \\
 &  \cite{benhamou2021detecting} & \cmark & \cmark  & \cmark & &  & & & & & & \cmark & & &  & &  \\
 &  \cite{liu2018practical}  &  \cmark & \cmark  & & &  & & & & & & & \cmark & &  & &  \\
 &  \cite{wang2019alphastock} & \cmark & & & & \cmark  & & & \cmark & & & \cmark & & &  & &  \\
 &  \cite{li2019optimistic}  &  \cmark & \cmark  & & &  & & & & & & \cmark & & &  & &  \\
 &  \cite{park2020intelligent}& \cmark & \cmark  & & &  & & & & & & & \cmark & &  & &  \\
 &  \cite{gao2020application} & \cmark & \cmark  & & &  & & & & & & \cmark & & &  & &  \\
 &  \cite{koratamaddi2021market} & \cmark & \cmark  & & &  & \cmark & & & & & & \cmark & &  & &  \\
 &  \cite{yang2020deep}&  \cmark & \cmark  & & \cmark  & & & & & & & \cmark & & &  & &  \\
 &  \cite{liu2020adaptive}&  \cmark & \cmark  & & \cmark  & & & & & & & & \cmark & &  & &  \\
 &  \cite{theate2021application} & \cmark & \cmark  & & &  & & & & & & & \cmark & &  & &  \\
 &  \cite{lei2020time} &  \cmark & \cmark  & & &  & \cmark & & & & & & & &  & &  \\
 &  \cite{ma2021parallel} &  \cmark & \cmark  & & \cmark  & & & & & & & & \cmark & &  & &  \\
 &  \cite{lee2020maps} &  \cmark & & & &  & & & & & & & \cmark & &  & &  \\
 &  \cite{soleymani2021deep} &  \cmark & \cmark  & & \cmark  & & & & & & & \cmark & & &  & &  \\ \hline
\multirow{8}{*}{\rotatebox{90}{Optimal Execution}}  & \cite{nevmyvaka2006reinforcement} &  & \cmark  & & &  & & \cmark  &  \cmark & & \cmark  &  & & \cmark  &  & &  \\
 &  \cite{deng2016deep}  &  \cmark & & & &  & & & & & & & \cmark & &  & &  \\
 &  \cite{wei2019model}  & & \cmark  & & &  & & & & \cmark &  & & & & \cmark & &  \\
 &  \cite{lin2020E2E} & & \cmark  & & &  & & & & \cmark &  & & & & \cmark & &  \\
 &  \cite{fang2021universal} & \cmark & \cmark  & & &  & & & \cmark & & \cmark  &  & & & \cmark & &  \\
 &  \cite{lin2020deep}&  \cmark & \cmark  & & &  & & \cmark  &  & \cmark &  & & & & \cmark & &  \\
 &  \cite{shen2014risk}  & & \cmark  & & &  & & \cmark  &  & & & & & \cmark  &  & &  \\
 &  \cite{hendricks2014reinforcement} &  & \cmark  & & &  & & \cmark  &  \cmark & & \cmark  &  & & & \cmark & &  \\ \hline
\multirow{9}{*}{\rotatebox{90}{Market Making}}& \cite{chan2001electronic} & \cmark & \cmark  & & &  & & \cmark  &  & & & & & &  & \cmark &  \\
 &  \cite{Patel2018MM} &  \cmark & \cmark  & & &  & & & \cmark & \cmark &  \cmark  &  & & &  & \cmark &  \\
 &  \cite{Spooner2018MM}  &  \cmark & \cmark  & & \cmark  & & & \cmark  &  \cmark & & & & & &  & \cmark & \cmark  \\
 &  \cite{Ganesh2019MM}&  \cmark & \cmark  & & &  & & \cmark  &  \cmark & & & & & &  & \cmark &  \\
 &  \cite{mani2019applications}  &  & & & &  & & \cmark  &  & & & & & &  & \cmark &  \\
 &  \cite{spooner2020robust} & & \cmark  & & &  & & & & & \cmark  &  & & &  & \cmark &  \\
 &  \cite{gavsperov2021market}& \cmark & \cmark  & & &  & \cmark & & & & & & & &  & \cmark &  \\
 &  \cite{selser2021optimal} &  \cmark & \cmark  & & &  & & & & & \cmark  &  & & &  & \cmark &  \\
 &  \cite{haider2021gaussian} & \cmark & \cmark  & & \cmark  & & & \cmark  &  & & & & & &  & \cmark & \cmark  \\ \hline
\end{tabular}
\label{table:state_action}
\end{table}

\begin{table}[H]
\caption{Literature on RL methods in finance: reward and algorithm. }
\label{table:reward+algo}
\vspace{.4cm}
\hspace*{-2.5cm}
\begin{tabular}{lp{0.3\textwidth}|lllllll|lllllllllll}
\hline
\multirow{2}{*}{} & \multirow{2}{*}{Literature}& \multicolumn{7}{l|}{Reward}& \multicolumn{9}{l}{Algorithm} &\\ \cline{3-20} 
  & & \rotatebox{90}{PnL} & \rotatebox{90}{Return} & \rotatebox{90}{VR} & \rotatebox{90}{IS} & \rotatebox{90}{Shaped reward} & \rotatebox{90}{Inventory Penalty} & \rotatebox{90}{Market Quality} & \rotatebox{90}{DQN} & \rotatebox{90}{PG} & \rotatebox{90}{DDPG} & \rotatebox{90}{A2C} & \rotatebox{90}{PPO} & \rotatebox{90}{simulator} & \rotatebox{90}{Q-learning} & \rotatebox{90}{SARSA} & \rotatebox{90}{NAC-S} & \rotatebox{90}{R-learning}& \rotatebox{90}{neuroevolution}\\ \hline
\multirow{20}{*}{\rotatebox{90}{Portfolio Management}} & \cite{jiang2017deep} & & \cmark & & & & &  & & \cmark & & & & & & &\\
  & \cite{yu2019model} & & \cmark & & & & &  & & & \cmark & & & & & &\\
  & \cite{zhang2020deep} & & & \cmark & & & &  & \cmark& \cmark & & \cmark& & & & &\\
  & \cite{ye2020reinforcement}  & & \cmark & & & & &  & & \cmark & & & & & & &\\
  & \cite{wang2021deeptrader} & & \cmark & & & & &  & & \cmark & & & & & & &\\
  & \cite{wang2021commission} & & \cmark & & & & &  & \cmark& \cmark & & & & & & &\\
  & \cite{benhamou2021detecting}& & \cmark & & & & &  & & \cmark & & & & & & &\\
  & \cite{liu2018practical} & & \cmark & & & & &  & & & \cmark & & & & & &\\
  & \cite{wang2019alphastock} & & & \cmark & & & &  & & \cmark & & & & & & &\\
  & \cite{li2019optimistic} & & \cmark & & & & &  & & & \cmark & & & & & & \\
  & \cite{park2020intelligent}  & & \cmark & & & & &  & \cmark& & & & & & & &\\
  & \cite{gao2020application} & & \cmark & & & & &  & \cmark& & & & & & & &\\
  & \cite{koratamaddi2021market}& & & & & \cmark & &  & & & \cmark & & & & & &\\
  & \cite{yang2020deep}  & & & & & \cmark & &  & & & \cmark & \cmark& \cmark& & & & \\
  & \cite{liu2020adaptive} & & & \cmark & & & &  & & \cmark & & & & & & &\\
  & \cite{theate2021application}& & \cmark & & & & &  & \cmark& & & & & & & &\\
  & \cite{lei2020time} & & \cmark & & & & &  & \cmark& & & & & & & &\\
  & \cite{ma2021parallel}& & \cmark & & & & &  & & \cmark & & & & & & &\\
  & \cite{lee2020maps} & & & & & \cmark & &  & \cmark& & & & & & & &\\
  & \cite{soleymani2021deep}& & & & \cmark & & &  & & & & \cmark& & & & &\\ \hline
\multirow{8}{*}{\rotatebox{90}{Optimal Execution}} 
& \cite{nevmyvaka2006reinforcement} & \cmark& & & & & &  & \cmark& & & & & & & &\\
  & \cite{deng2016deep} & \cmark& & \cmark & & & &  & & \cmark & & & & & & &\\
  & \cite{wei2019model} & \cmark& & & & & &  & & & & & & \cmark & & &\\
  & \cite{lin2020E2E} & & & & \cmark & & &  & & & & & \cmark& & & &\\
  & \cite{fang2021universal} & & & & & \cmark & &  & & & & & \cmark& & & &\\
  & \cite{lin2020deep}  & & & & & \cmark & &  & \cmark& & & & & & & &\\
  & \cite{shen2014risk} & \cmark& & & & & &  & \cmark& & & & & & & &\\
  & \cite{hendricks2014reinforcement} & & & & \cmark & & &  & \cmark& & & & & & & &\\ \hline
\multirow{9}{*}{\rotatebox{90}{Market Making}}  & \cite{chan2001electronic} & \cmark& & & & & \cmark & \cmark  & & & & & & & & \cmark & & \\
  & \cite{Patel2018MM} & \cmark& & & & & &  &\cmark  & & & & & & & & &  \\
  & \cite{Spooner2018MM} & \cmark& & & & & \cmark &  & & & & & & & \cmark & \cmark& &  \cmark \\
  & \cite{Ganesh2019MM}  & \cmark& & & & & \cmark &  & & & & & \cmark & & & &\\
  & \cite{mani2019applications} & \cmark& & & & & \cmark & \cmark  & & & & & & & & \cmark&  & \\
  & \cite{spooner2020robust}& \cmark& & & & & \cmark &  & & & & & & & & & \cmark & \\
  & \cite{gavsperov2021market}  & \cmark& & & & & \cmark &  & & & & & & & & & & &\cmark \\
  & \cite{selser2021optimal}& \cmark& & & & & &  & & & & & & & \cmark & & &  \\
  & \cite{haider2021gaussian} & \cmark& & & & & \cmark &  & & & & & & & & \cmark & & \\ \hline
\end{tabular}
\label{table:reward_algo}
\end{table}

%% file: Tables/tab_simulator_MM.tex
\begin{table}[]
    \centering
\begin{tabular}{m{2cm}  m{5 cm} m{7cm}} 
  \hline
  Literature & Security Mid-price Model & Investor Agent Model \\
  \hline

  \citet{chan2001electronic} & Poisson Jump Process &   Informed and uninformed trading
agent arrivals modeled by Poisson Process with constant intensity 
\\
  \hline
  \citet{mani2019applications} & Poisson Jump Process &   Informed and uninformed trading
agent arrivals modeled by Poisson Process with constant intensity
\\
  \hline

  \citet{selser2021optimal} & Brownian motion &   Investor agent orders modeled by Poisson processes with intensities being functions of the offset between the security price and the prices offered by the market maker \\
  \hline

  \citet{spooner2020robust} & Brownian motion with drift; drift parameter are not fixed but decided by adversary agent&   Investor agent orders modeled by Poisson processes with intensities being functions of the offset between the security price and the prices offered by the market maker; parameters of the function are not fixed but decided by adversary agent \\
  \hline

  \citet{Ganesh2019MM} & Geometric Brownian motion &   Investor agent orders are generated with a fixed probability at each timestep \\
  \hline

\end{tabular}
    \\[10pt]
    \caption{Simulators of market making with RL}
    \label{table:market_making_simulators}
\end{table}